# Deterministic polarization chaos in a laser diode

Martin Virte[1, 2, †], Krassimir Panajotov[2, 3], Hugo Thienpont[2] and Marc Sciamanna[1]

[1], Supelec OPTEL Research Group, Laboratoire Matériaux Optiques, Photoniques et Systèmes (LMOPS) EA-4423, 2 rue Edouard Belin, 57070 Metz, France
[2], Brussels Photonic Team, Department of Applied Physics and Photonics (B-PHOT TONA), Vrije Universiteit Brussels, Pleinlaan 2, 1050 Brussels, Belgium
[3], Institute of Solid-state Physics, 72 Tzarigradsko Chaussee Blvd., 1784 Sofia, Bulgaria
[†], marc.sciamanna@supelec.fr
[‡], martin.virte@supelec.fr

**Fifty years after the invention of the laser diode and fourty years after the report of the butterfly effect - i.e. the unpredictability of deterministic chaos, it is said that a laser diode behaves like a damped nonlinear oscillator. Hence no chaos can be generated unless with additional forcing or parameter modulation. Here we report the first counter-example of a free-running laser diode generating chaos. The underlying physics is a nonlinear coupling between two elliptically polarized modes in a vertical-cavity surface-emitting laser. We identify chaos in experimental time-series and show theoretically the bifurcations leading to single- and double-scroll attractors with characteristics similar to Lorenz chaos. The reported polarization chaos resembles at first sight a noise-driven mode hopping but shows opposite statistical properties. Our findings open up new research areas that combine the high speed performances of microcavity lasers with controllable and integrated sources of optical chaos.**

The discovery of deterministic chaos - i.e. the aperiodic deterministic dynamics of a nonlinear system showing sensitivity to initial conditions - has been a major paradigm shift overthrowing two centuries of Laplacian viewpoint of dynamical systems[1-5]. Looking into a system behavior with the use of chaos theory has helped to interpret and control many of such ordered or disordered behaviors in our present day life, such as the bifurcations leading to epilepsy and cancer[6], the stabilization of cardiac arrhythmias[7], and the improvement of complex behavioral patterns in robotics[8].

Soon after the invention of the laser, the possibility to observe light chaos raised attention. In 1975, Haken discovered an analogy between the Maxwell-Bloch equations for lasers and the Lorenz equations showing chaos[9]. The Maxwell-Bloch equations are three equations for the field E, the polarization P and the carrier inversion N, each with its own relaxation time. However, while in Lorenz equations the relaxation times of the dynamical variables are of similar order of magnitude, they may take very different values in lasers. If one variable relaxes much faster than the others, this variable is adiabatically eliminated, hence resulting in a reduced number of dynamical equations. Therefore, so-called class A (ex: He-Ne, Ar and Dye), class B (ex: Nd:YAG, $CO_2$ and semiconductor) or class C (ex: $NH_3$) lasers have dynamics governed either by a single equation for the field, two equations for the field and population inversion or the full set of equations, respectively. In class A or class B laser systems chaos cannot be observed unless one adds one or several independent control parameters[10]. Chaos has then been reported in, for example, free-running $NH_3$ lasers[11], He-Ne lasers with modulation of the external field[12], $CO_2$ lasers with loss modulation[13], solid-state laser with gain modulation[14], injected field[15] or global multimode coupling[16], diode lasers with optical feedback[17], saturable absorption[18], or optical injection[19].

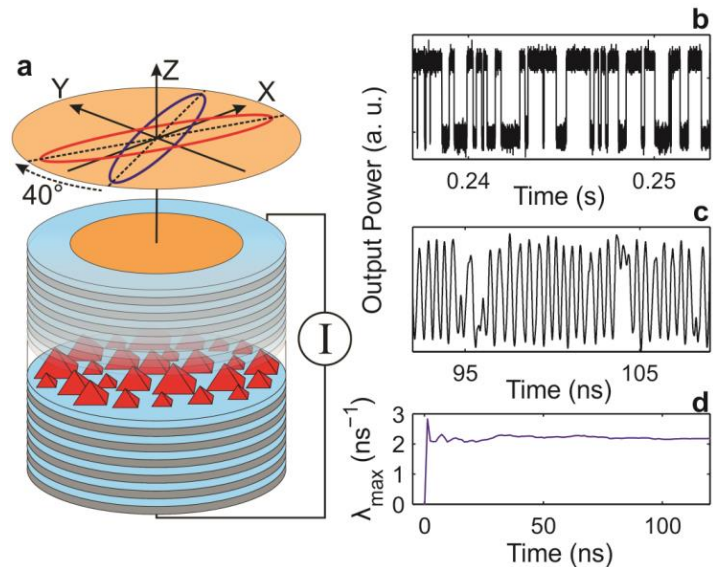

*Figure 1:* ***Experimental observations showing chaotic polarization mode hopping in a quantum-dot VCSEL.*** *(a) Artist view of a Quantum Dot VCSEL with multiple quantum dots confined between two mirrors. At threshold the laser emits X linearly polarized light; Driven by constant injection current, the device can exhibit chaotic mode hopping between two non-orthogonal elliptically polarized modes (red and blue ellipses) separated by about 40 degrees. (b-c) Polarization resolved output power time trace showing the chaotic polarization mode hopping at a constant injection current of 2.0 and 2.6 mA respectively. (d) Estimation of Largest Lyapunov exponent for the time-series in (c).*

In this paper we demonstrate a paradigm shifting phenomenon: deterministic chaos in a free-running laser diode, i.e. without external forcing or modulation. Chaos is unambiguously identified in the light polarization output of a vertical-cavity surface-emitting laser (VCSEL).
Moreover, both the route to chaos and the mode dwell time – time the laser stays emitting in one mode - characteristics agree qualitatively very well with the longstanding and so far unconfirmed prediction of the spin-flip model[20]: a nonlinear coupling between two polarization modes induced by carrier spin relaxation leads to deterministic nonlinear dynamics.
Today applications of laser diodes such as secure communications[21-23] and random number generation[24,25] make use of high-dimensional chaos obtained from time-delayed optical feedback. The chaos however shows correlation at the time-delay[26-28], hence making it easy to reconstruct the attractor in a low dimensional phase space and therefore reducing both security and randomness. The polarization chaos reported here is of low-dimension but (1) is obtained from a free-running device, i.e. without the additional complexity of optical feedback, and (2) uncovers a complex polarization dynamics that may improve randomness[29] and security[30,31] and that allows for chaos multiplexing at high speed[32].

**Route to polarization chaos**

Figure 1 shows typical measured dynamics (b-c) of the polarization-resolved laser output power in a single-mode submonolayer 990-nm quantum dot VCSEL (a). The device characteristics, in particular its polarization and spectral properties as a function of the injection current, have been detailed elsewhere[33]. When increasing the current, the VCSEL switches from lasing with a linearly polarized emission to lasing in one of two elliptically polarized modes. The main axes of the ellipses make an angle of 40 degrees; hence the two polarization modes are not orthogonal. In (b) is shown the polarization resolved output measured at +45 degrees according to the linear eigenaxes (X and Y axes of figure 1-(a)). The laser exhibits a polarization dynamic that resembles a random-like hopping between two polarization modes. At first sight it looks very similar to the so-called stochastic polarization mode hopping that has been extensively reported in VCSELs showing current or temperature driven polarization bistability[34]. However this polarization dynamic shows many features that are strikingly different from the stochastic mode hopping: (1) as schematically plotted in (a) the competing polarization modes are elliptically polarized and not orthogonal, (2) a spectral analysis of the polarization-resolved output unveils an underlying self-pulsating dynamics at about 8 GHz, which corresponds to undamped relaxation oscillations, (3) the time-scale of the dynamics strongly depends on the injection current, see (b) and (c): the mean-dwell-time (time the laser stays emitting in one mode) decreases with the increase of current and changes over several orders of magnitude - from seconds to nanoseconds - in a small current range[33] and (4) using Wolf's algorithm[35], we find a positive maximum Lyapunov exponent from the experimental time series. An example is shown in figure 1 (d) for the dynamics shown in (c), but a similar conclusion holds for different values of the injection current and therefore for either slow (s) or fast (ns) mode hopping dynamics.

The same polarization properties are seen in seven out of twenty-six devices, hence they are not singular and can be easily reproduced. These polarization characteristics call for an interpretation that is different from a stochastic Kramers hopping problem applied to symmetric bistable states, as is typically done for VCSELs[34]. Our experiment can qualitatively be reproduced well using the San Miguel, Feng and Moloney (SFM) model. The model considers two competing emission processes for the right and left circular polarization with two coupled carrier reservoirs[20,36]; the coupling comes from multiple complex spin-flip processes, i.e. sub-picosecond mechanisms that contribute to the relaxation of electron spin (see e.g. Ref.[37]), modeled as a single average decay rate $\gamma_s$ for simplicity. The resulting six-dimensional set of rate equations explains polarization switching in VCSELs as resulting from a cascade of bifurcations on linearly polarized steady-states, which appear when increasing the injection current. Although nonlinear dynamics have been suggested in experiments[33,38], no evidence of a chaotic behavior of a free-running VCSEL has been provided so far.

To do so we make direct numerical integrations of the SFM model using the following parameters (notations are identical to the one used by Martin-Regalado et al.[36]): amplitude anisotropy $\gamma_a = -0.7$ ns$^{-1}$, phase anisotropy $\gamma_p = 4$ ns$^{-1}$, linewidth enhancement factor $\alpha = 3$, spin-flip processes decay rate $\gamma_s = 100$ ns$^{-1}$, decay rate of the electric field in the cavity $\kappa = 600$ ns$^{-1}$, decay rate of the total carrier number $\gamma = 1$ ns$^{-1}$ and normalized injection current $\mu$ in [1, 10]. We use a classic 4-stage Runge Kutta algorithm with a time step of 1 ps.

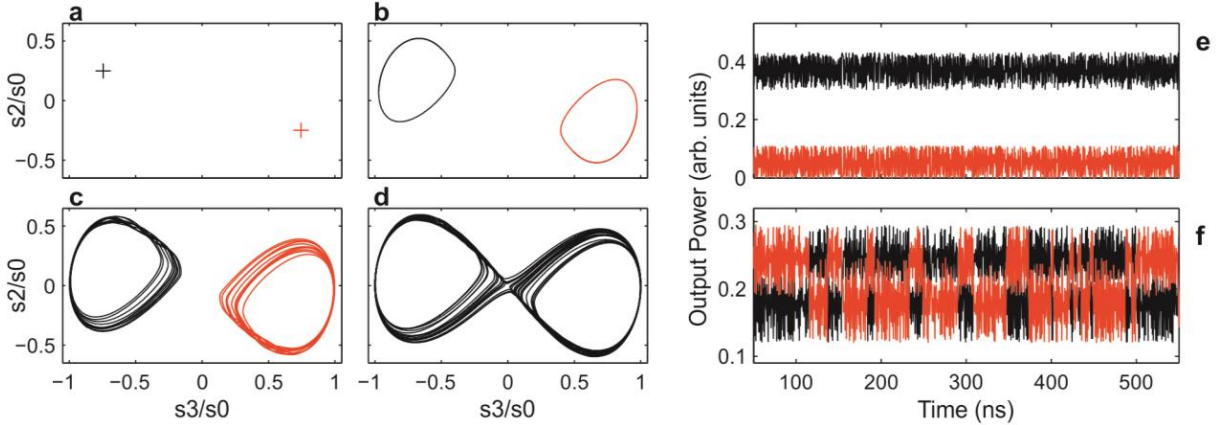

*Figure 2: **Route to chaos in SFM framework.** Bifurcation sequence in Stokes parameter (s3; s2) plane for increasing injection current µ. (a) µ = 1:34, two symmetric non-orthogonal elliptically polarized (EP) states. (b) µ = 1:39, two limit cycles oscillate around unstable EP states. (c) µ = 1:41, cascade of period doubling bifurcations leads to single-scroll chaotic attractors. (d) µ = 1:425, the two attractors merge into a single double-scroll attractor. (e-f) Polarization resolved output power at µ = 1:425: (e) projections at 0 (black) and 90 degrees (red) and (f) projections at 45 (black) and -45 degrees (red).*

At threshold the laser emits linearly polarized light but when the injection current is increased, it experiences a complex bifurcation sequence leading to chaos. This route to polarization chaos is explained in figure 2 where we display the evolution of the system trajectory in the (s3; s2) plane for different current values; the $s_x$ being the Stokes parameters. At threshold, the laser operates in a linearly polarized steady-state. As the current is increased, this state is destabilized by a pitchfork bifurcation creating two symmetric elliptically polarized states: they are displayed as the black and red crosses in figure 2-(a). When the current is increased, the polarization rotates and the ellipticity of the output beam increases. The two elliptically polarized states are never orthogonal for any given injection current. Then if we keep increasing the current, both elliptically polarized states become unstable and two limit cycles are created, oscillating around these unstable steady-states, as shown in figure 2-(b). Soon after a sequence of period-doubling bifurcations occurs and two symmetric single-scroll chaotic attractors appear, see figure 2-(c). The trajectories of both attractors are centered on the unstable elliptically polarized states, and the size of the attractors grows as the current is increased. Therefore beyond a critical value of the injection current they merge into a single double-scroll attractor which reminds the "butterfly" attractor of Lorenz chaos (see figure 2-(d)).This chaotic behavior, obtained by simulation, is displayed in figure 2-(e, f) where we plot the polarization resolved output powers at an injection current of µ = 1:425 for different projections. All time traces are averaged over 0:5 ns to simulate a slower photodetector response. In figure 2-(e) we plot the projections at 0 and 90 degrees, i.e. the polarization at threshold and the orthogonal one. We clearly observe a "noisy" but non-switching operation of the laser with an emission in both polarizations, however for different projections we find a completely different behavior. Indeed, in figure 2-(f) the output power for projections at 45 and -45 degrees displays irregular, random-like switching between two different states as observed in the experiment. At this point it is worth mentioning that the system remains chaotic at all times (with at least one positive Lyapunov exponent) even when it stays in one of two elliptically polarized states, i.e. when the trajectory remains on one wing of the double scroll butterfly attractor. Numerically we also were able to demonstrate that: (1) a similar chaotic polarization hopping is observed in a large range of laser parameters and (2) the self-pulsation from where chaos originates has a frequency that is either close to the birefringence induced frequency splitting or to the relaxation oscillation frequency. The laser's dynamics therefore shows pulsations at a frequency of typically several GHz.

**Statistics of the deterministic mode hopping**

We further analyze the statistics of the dwell time, i.e. the time the laser stays in one polarization state, and how the averaged dwell time scales with the laser injection current. Considering the similarity between the chaotic polarization mode hopping and the Lorenz chaotic attractor, we also simulate the case of the Lorenz chaos equations. In Lorenz's equations, the r parameter plays the same role as the injection current in a laser[3,9]. The results are displayed in figure 3: (a, b) for the laser system and (c, d) for Lorenz chaos. In (a) and (c) we give the evolution of the dwell-time for increasing the injection current and the r parameter respectively. The dwell time distributions are given in (b) and (d) for injection current of $\mu = 1{:}425$ and r parameter of 27.

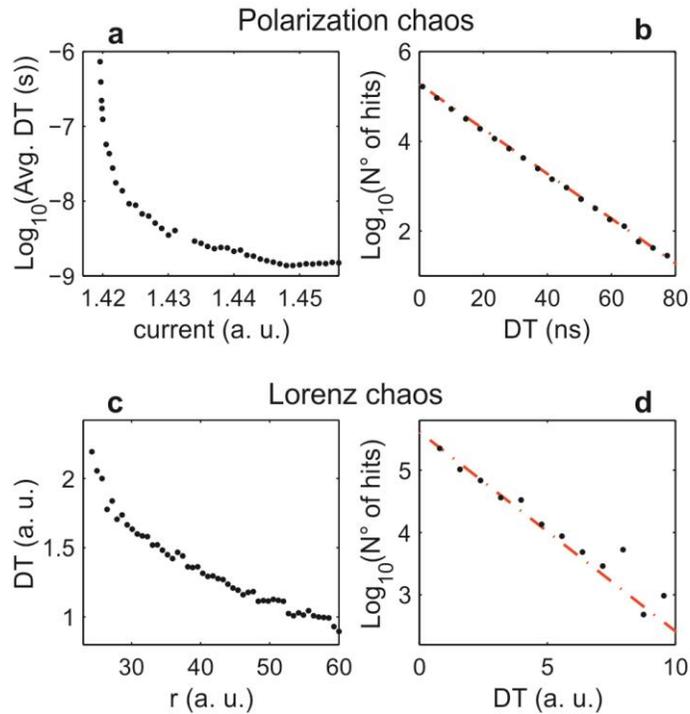

*Figure 3: **Statistical properties of mode dwell time and comparison between polarization and Lorenz chaos.** Dwell-time (DT) statistics for simulated polarization chaos in laser diode (a, b) and for Lorenz chaos (c, d). (a) Semi-Logarithmic plot of average dwell time for polarization chaos versus injection. (b) Distribution of the dwell time for polarization chaos at $\mu = 1{:}425$. (c) Average dwell time evolution for Lorenz chaos versus r parameter. (d) Distribution of the dwell time for Lorenz chaos with $r = 27$.*

Figure 3-(a) shows a significant decrease of the dwell time with current: as the current is increased, the dwell time scales from microseconds to nanoseconds. The dwell time statistical distribution follows an exponential decay law, as shown in Fig. 3-(b). These two statistical features agree qualitatively well with what is seen in our experiment[33]. It is worth mentioning that the exponentially decaying statistical distribution of the dwell time is the result of polarization nonlinear deterministic dynamics and, in contrast to all previous studies, is not the result of a noise-driven Kramers hopping problem. In the latter case, polarization mode hoping between linearly polarized VCSEL modes displays completely opposite behavior: it exponentially increases with the injection current[34].

The Kramer's approach therefore fails to characterize the dwell time statistics of figure 3. Since the time series shows a random-like switching between two states it is of interest to check whether the dynamics can be alternatively modeled as a hidden markov process[39,40], i.e. to analyze whether the jumps between states are influenced by hidden additional variables or any external perturbation. Such a modeling would then allow us to identify the most probable noise-induced transitions. To do so, we apply the Baum and Welch algorithm to retrieve two matrices: the transition Markov matrix (M) and the hidden-related transition matrix (N). In the case of the experimental time series (Fig 1 (c)) we obtain off-diagonal elements of N very close to zero (less than $10^{-10}$), which means that the random-like jumps are not the result of noise, but of the internal system dynamics[40]. The same result that confirms the deterministic chaos as a driving force for the polarization mode hopping and the random-like switching is also found from simulations on the SFM model, even in presence of laser spontaneous emission noise, as well as on the Lorenz equations.

In summary, the similarities between the VCSEL polarization chaos experiments and simulations and the Lorenz chaos simulations, in addition to the fact that both Kramer's and hidden markov approaches fail to explain the mode hopping statistics, lead us to the conclusion that the physical mechanism underlying the mode hopping dynamics and statistics is a chaotic trajectory of the system in a double scroll attractor. When increasing r or the injection current, the two wings of the butterfly chaotic attractor grow, hence making the jumps between the two wings easier and therefore reducing the dwell time.

**Experimental chaos identification**

In this section, we clearly demonstrate that the reported experimental dynamic is chaotic. Here we use the Grassberger-Procaccia (GP) algorithm which gives an estimate of the correlation dimension $D_2$ and of the $K_2$-entropy (Kolmogorov-entropy)[41]: $K_2$ is zero for periodic or quasiperiodic systems, positive for chaos and $K_2 = 1$ for purely stochastic processes. Considering a sequence of N points, we divide it into N-D vectors of size D and we compute the correlation integral $C_D(r)$ for this new sequence, i.e. the average number of vectors that can be found in a sphere of radius r around a single vector. If it converges, the slope of $Log(C_D(r))$ should give us an estimation of the correlation dimension $D_2$; it also allows us to estimate the $K_2$-entropy, see Ref.[41]. To improve the efficiency of the GP algorithm on noisy experimental time-series we also use an extra re-embedding procedure based on singular-value analysis[42].

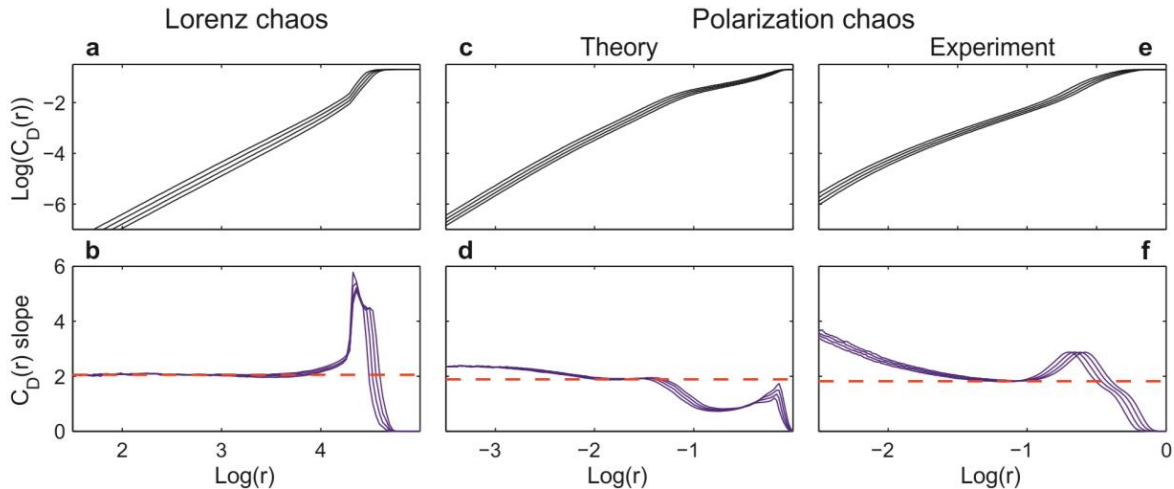

*Figure 4: **Chaos identification.** Results of the Grassberger-Procaccia algorithm for Lorenz chaos (a-b), theoretical polarization chaos (c-d) and experimental time-series (e,f). (a), (c) and (e) logarithmic plots of correlation integral $C_D(r)$ versus sphere radius r. (b), (d) and (f) slope of the correlation integral versus sphere radius r. In the three cases the GP algorithm converges creating a plateau on the slope of the correlation integral, the red dashed curves give estimations of the correlation dimension in all three cases: Lorenz chaos D2 ~ 2:05, theoretical polarization chaos D2 ~ 1:89 and experimental time-series D2 ~ 1:82.*

Results of the GP algorithm are displayed in figure 4 for Lorenz chaos, theoretical polarization chaos and experimental time-series in figure 4-(a,b), (c,d) and (e,f) respectively. Figures 4-(a), (c) and (e) are logarithmic plots of the correlation integral versus sphere radius while figures 4-(b), (d) and (f) are plots of the correlation integral slope versus sphere radius. In all three cases we obtain a clear convergence toward a correlation dimension of $D_2 \sim 2$: for each time-series, a well-resolved plateau can be seen in the plot of the correlation integral slope. We obtain the following values for the $K_2$ entropy: $K_2 \sim 1.65$ for Lorenz chaos, $K_2 \sim 2.5$ for polarization chaos and $K_2 \sim 7.1$ for experimental data. On the other hand, we verified that no convergence appeared in a case of a stochastic polarization mode hopping, as can be simulated from the noise-driven two-mode equations (1)- (3) of Ref.[43]. As expected for a noise induced process, $C_D(r)$ keeps increasing along with D and never converges for any r, which would then lead to an infinite value of $K_2$.

Because specific colored noise can also exhibit finite non-zero $K_2$-entropy, few additional tests suggested by Provenzale et al.[44] are performed to fully confirm the chaotic behavior of our dynamics. Our experimental time-series passed all these tests successfully and we only report one of them which is complementary to the GP algorithm, easy to implement and with simple and visual results.

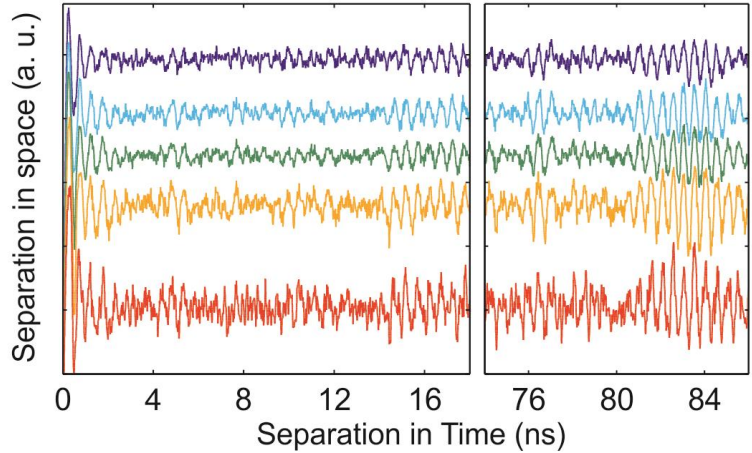

*Figure 5: **Discrimination between chaos and colored noise.** We plot a space-time separation contour map for the experimental time-series. The different curves correspond to different fraction of the distribution: above the blue line is 10% of the points, then going from blue to red the fractions are 30, 50, 70 and finally 90% of points are above the red curve. The dropouts indicate a particular proximity in space between pairs of point for a given time-shift. Such dropouts can be found in the whole range of time-shift we consider, i.e. several tens of nanoseconds, demonstrating a strong dynamical correlation.*

For noise driven processes two points will be correlated only if they are close in time; on the contrary for chaotic processes a larger correlation exists between points as the system is driven by a low-dimensional attractor: order can be found in the random-like dynamic. As the GP algorithm does not consider the time-separation between points, specific noises can produce a finite $K_2$ value. In figure 5, we plot the separation in space versus the separation in time for a large number of pairs of points chosen randomly in our time-series. Noise induced processes will exhibit stationary distribution after a small transient as they are not correlated. Instead we find frequent and sharp dropouts. These dropouts are manyfold, irregularly distributed and appear in the whole range of time-shifts we consider, i.e. several tens of nanoseconds. This result therefore confirms that our time-series are not noise driven.

**Discussion**

In summary, we report on the first observation of a deterministic chaos in the dynamics of a free-running laser diode. Chaos results from the combination of (1) the light polarization degree of freedom in a vertical-cavity surface-emitting laser and (2) nonlinear coupling mechanisms between two lasing modes with elliptical polarizations, as modeled within the so-called San Miguel, Feng and Moloney (SFM) approach for polarization switching. Increasing the injection current leads to polarization chaotic two mode hopping similar to the trajectories of the Lorenz chaos when jumping between the two wings of its double scroll attractor. The chaotic property of the experimental and numerical time-traces is demonstrated using appropriate chaos identification techniques.

We anticipate our findings to have an impact in several directions. First, the reported polarization chaos has dwell time hopping properties opposite to those of the so far reported noise-driven mode hopping. Which material and device properties lead to either chaos polarization switching or stochastic polarization switching is a challenging question in the development of polarization-controlled microcavity lasers. Since the underlying physics is a nonlinear carrier spin coupling, one could take advantage of and make a connection with recent works on spin control by carrier injection[45] or variation of the quantum well composition[46]. Secondly, the polarization chaos reported here has frequencies that are not limited by the laser relaxation oscillations, as suggested by the SFM model and as can be expected from the recent works on spin-controlled VCSEL[47]. Our findings therefore pave the way towards the development of integrated microcavity lasers generating multi-gigahertz chaos. Finally, polarization chaos encoding has only been used once in an experiment showing message encoding and decoding at about 100 Mb/s using synchronized polarization fluctuations in birefringence-modulated erbium doped fiber ring lasers[48]. Our work suggests the development of polarization chaos communication at higher modulation rates - beyond 10 GHz - and with compact, low-cost and low threshold microcavity lasers. Today state-of-the-art high bit rate optical communications indeed make use of polarization multiplexing (in combination with phase encoding and coherent detection) and have overcome many of the limitations due to polarization mode dispersion[49]. Integration of polarization encoding/decoding is therefore relevant. Polarization chaos also shows many advantageous properties for chaos multiplexing[32], chaos synchronization and security[31] and ordering in globally coupled oscillators[50] which now can be explored and practically tackled.

**Methods**

**Simulation details.** For the SFM model, we use a classic 4 step Runge-Kutta algorithm with a time step of 1 ps. For the application of the Grassberger-Procacia algorithm we only consider the X polarization output power. For the Lorenz's equations, we use again a classic 4 step Runge-Kutta algorithm with a time step of 0.004. For the application of the Grassberger-Procacia algorithm we only consider the y variable.

**Statistical analysis.** Evolution of the dwell-time for polarization chaos is measured over 2000 mode-hopping events. For Lorenz chaos, we measured the dwell-time evolution over 10000 events. Both distributions have been calculated for 500,000 mode-hopping events.

**Experimental data and re-embedding procedure.** Our main contribution is the analysis of experimental data. Measurements of the data are already described elsewhere[33] and we will therefore focus on the method to reduce the noise in our dataset. Here we use a re-embedding procedure based on singular value analysis as described by Fraedrich[42]. We shall keep the same notations.
Experimental observations are made with a Tektronix CSA 7404 digital oscilloscope (optical bandwidth channel of 4 GHz - sample rate of 20GS/S - used with a 2.4 GHz bandwidth photodiode) and we use time-series of 200 ns length. We use a window of size M = 0.6 ns which is about a dozen times the time-step of the measurements and finally we keep the 5 principal components of our time-series using the same procedure than the one described in Ref.[42].

**Chaos identification**: To identify the chaotic dynamic and to find order in experimental and theoretical time-series, we use the Grassberger-Procacia (GP) algorithm[41]. The goal of this algorithm is to provide an estimation of the correlation dimension $D_2$ (close to the fractal dimension of the attractor) and the $K_2$-entropy which characterizes the randomness of a dynamic (close to the Kolmogorov Entropy). We use the same parameters for the Grassbeger-Procacia algorithm in all three cases. The N points sequence is divided into N-D vectors of size D and for each value of D we compute the correlation integral $C_D(r)$:

$$C_D(r) = \frac{1}{N^2}\left(\sum_{n,m}(d(X_n, X_m) < r)\right)$$

This function is a numerical computation of the average number of vectors that can be found within a sphere of radius r around a given vector. The distance d is the Euclidian norm or norm 2.
According to Grassberger and Procacia[41], we theoretically have: $C_D(r) \sim r^\nu \exp(-\tau D \kappa)$, with $\tau$ the sample rate of the time-series; $\nu$ and $\kappa$ being the two parameters we want to approach. Thus we find:

$$D_2 = \lim_{\substack{D\to\infty \\ r\to 0}} \frac{d\ln(C_D(r))}{d\ln(r)} \text{ and } K_2 = \lim_{\substack{D\to\infty \\ r\to 0}} \frac{1}{\tau}\ln\left(\frac{C_D(r)}{C_{D+1}(r)}\right).$$

Obviously infinite is numerically out of range, therefore we look for convergence in these functions as we increase the vector size D. In this contribution we present results for D between 12 and 15.

**Discrimination between chaos and colored noise.** The goal of this method is to take explicitly in to account the time correlation between points, see Ref.[42].
This discrimination is based on a space-time separation plot. For a pair of vectors (Xn, Xm), we consider the separation in space d(Xn, Xm) and in time: Δt = |n-m|. To obtain a regular distribution in time, for each given value of Δt we select randomly in our experimental dataset 500 pairs of points. Then we plot the contour map which gives us the evolution of this distribution for increasing time-shift Δt.

**Acknowledgments**
The authors acknowledge the support of Conseil Régional de Lorraine, Fondation Supelec, FWO-Vlaanderen, METHUSALEM program of the Flemish government, and IAP VII 'Photonics@br' research program of the Belgian Federal Government.

**Author contribution:** M.S. & K.P. initiated the study. M.V. and M.S. performed the simulation of the laser dynamics. M.V. performed the chaos identification from experimental and theoretical time-traces. All authors discussed the results and wrote on the manuscript.

**Author Information:** Reprints and permissions information is available at www.nature.com/reprints. Authors declare no competing financial interest. Correspondence and requests for additional materials should be addressed to M.S.